\documentclass[aps,prl,superscriptaddress,floatfix,showpacs,notitlepage,twocolumn]{revtex4-1}
\usepackage[latin9]{inputenc}
\usepackage{color}
\usepackage{textcomp}
\usepackage{amsmath}
\usepackage{amssymb}
\usepackage{graphicx}
\usepackage{esint}
\usepackage{framed} 
\usepackage{xcolor}
\usepackage{hyperref}
\usepackage{footmisc}
\makeatletter
\usepackage{amsfonts}
\usepackage{bbold}
\usepackage{natbib}
\usepackage{comment}
\pdfoutput=1
\newcommand{\ket}[1]{|#1 \rangle}
\newcommand{\bra}[1]{\langle#1 |}

\newcommand{\lr}[1]{\left( #1 \right)}

\newcommand{\no}{\nonumber}

\newcommand{\mc}[1]{\mathcal{#1}}
\newcommand{\Q}{$\mathcal{Q}$ }

\begin{document}

\title{Photonic tensor networks produced by a single quantum emitter}

\author{Hannes Pichler}
\email[These authors contributed equally to this work\\Corresponding author: ]{hannes.pichler@cfa.harvard.edu}
\affiliation{ITAMP, Harvard-Smithsonian Center for Astrophysics, Cambridge, MA 02138, USA}
\affiliation{Physics Department, Harvard University, Cambridge, Massachusetts 02138, USA}

\author{Soonwon Choi}
\email[These authors contributed equally to this work\\Corresponding author: ]{hannes.pichler@cfa.harvard.edu}
\affiliation{Physics Department, Harvard University, Cambridge, Massachusetts 02138, USA}

\author{Peter Zoller}
\affiliation{Institute for Theoretical Physics, University of Innsbruck, A-6020
Innsbruck, Austria}
\affiliation{Institute for Quantum Optics and Quantum Information of the Austrian
Academy of Sciences, A-6020 Innsbruck, Austria}

\author{Mikhail D. Lukin}

\affiliation{Physics Department, Harvard University, Cambridge, Massachusetts 02138, USA}

\begin{abstract}
We propose and analyze  a protocol to generate two dimensional tensor network states using a single quantum system that sequentially interacts with a 1D string of qubits. This is accomplished by using parts of the string itself as a quantum queue memory. As a physical implementation, we consider  a single atom or atom like system coupled to a 1D waveguide with a distant mirror, where guided photons represent the qubits while the mirror allows the implementation of the queue memory. We identify the class of many-body quantum states that can be produced using this approach. These  include universal resources for measurement based quantum computation and states associated with topologically ordered phases. We discuss an explicit protocol to deterministically create a 2D cluster state in a quantum nanophotonic experiment, that allows for a realization of a quantum computer using a single atom coupled to light.
\end{abstract}

\date{\today}

\maketitle

Controlled generation of multi-qubit entanglement is central to quantum information science \cite{Kimble:2008if,Gisin:2007by,Briegel:2009gg}. In particular, quantum communication requires the use of photonic qubits, 
where information is encoded in the photon number or the polarization degrees of freedom of light. 
By coupling a single atom or atom-like system to a photonic waveguide, one can deterministically produce photonic qubits as well as atom-photon entanglement. Indeed, a systematic control over such single quantum emitters has been demonstrated in a variety of experimental systems  \cite{Mitsch:2014fz,Goban:2014eq,Sollner:2015fc,Sipahigil:2016hy,vanLoo:2013df,Hoi:2015fh,Tiecke:vw,Reiserer:2014hf,He:2013bq,Lodahl:2015fy} and used for fundamental tests of quantum mechanics \cite{Hensen:2015dw}.
More generally, it is known that combining quantum emitters with minimal resources such as quantum memories and few qubit registers can provide a powerful platform for quantum networks \cite{Kimble:2008if,Duan:2001dt,Reiserer:2015en}.

Beyond quantum communication applications, 
recently it has been proposed and demonstrated that individual quantum emitters  can be used to produce a sequence of photons that are entangled in a multipartite way \cite{Schon:2005fk,Lindner:2009eu,Economou:2010jt,Schwartz:2016dj}. 
This can be potentially of interest for quantum computation and simulation \cite{Osborne:2010hr,Barrett:2013ef,Eichler:2015if}. However, the entanglement structure of the resultant many-body state is characterized by so called matrix product states (MPS) \cite{Schon:2007bw,Verstraete:2010bf,Schollwock:2011gl}, which can be efficiently simulated classically, hence limiting their potential utility.    

This Letter describes a method to generate quantum states with higher dimensional entanglement structures, using minimal resources. 
The key idea is to employ a quantum memory (such as a delay line for photons) that allows for repeated interactions 
between a small quantum system (such as a quantum emitter) with a 1D string of individual qubits  \cite{Pichler:2016bx,Grimsmo:2015gf}. 
Specifically, we describe  an explicit protocol to create the 2D cluster state \cite{Rausssendorf:2001js}, a universal resource for quantum computation \cite{Raussendorf:2003ca} using photonic qubits interacting with a single few-state quantum system. More generally, we characterize the class of states achievable in this setting in terms of so-called projected entangled pair states (PEPS) \cite{Verstraete:2004cf}. This opens new avenues for quantum information processing \cite{Rausssendorf:2001js} and photonic simulation of quantum many-body physics with currently available experimental techniques \cite{Osborne:2010hr,Barrett:2013ef,Eichler:2015if}.

\begin{figure}[b]
\includegraphics[width=0.5\textwidth]{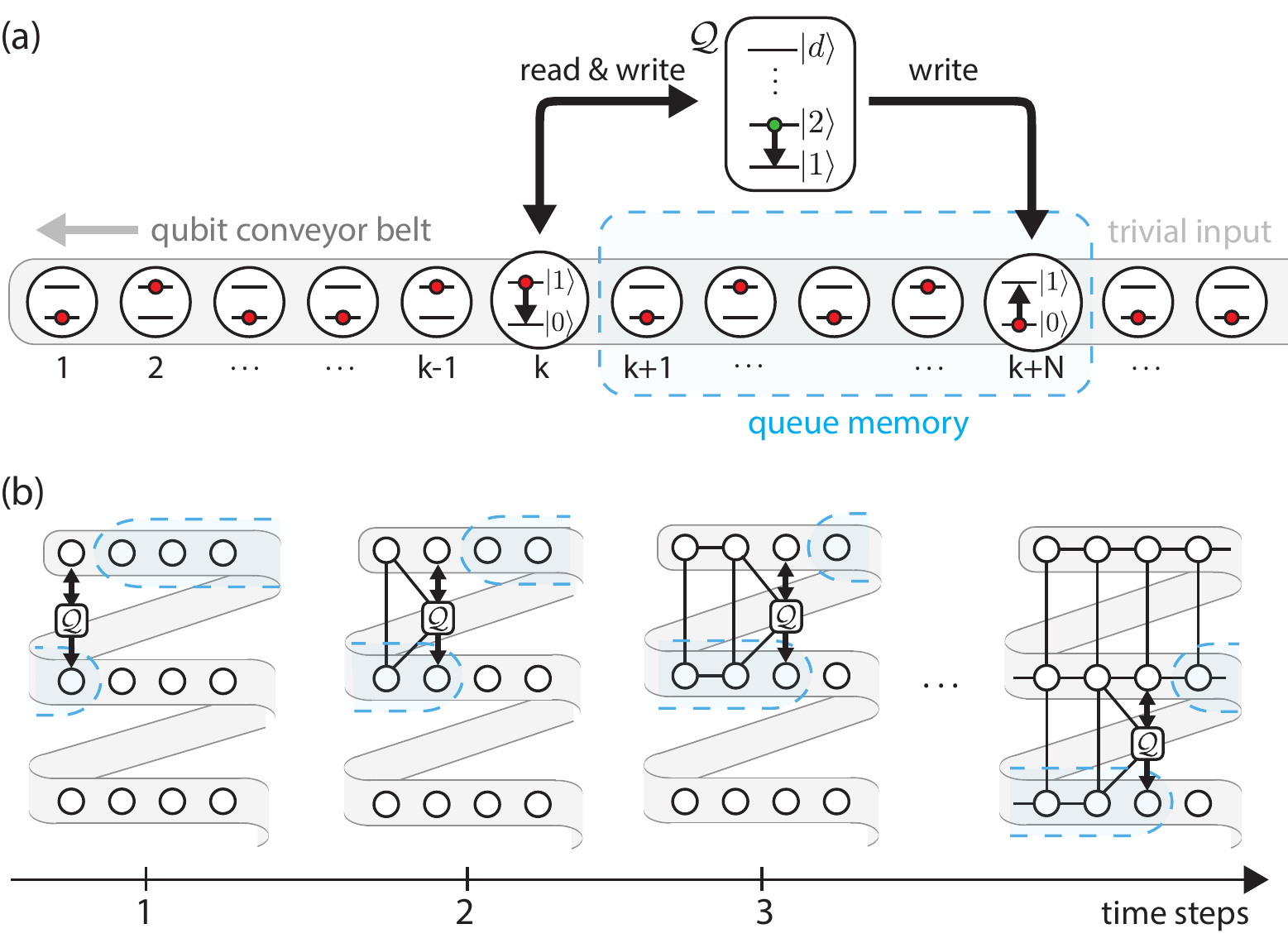}
\caption{Schematic setting. (a) We consider a $d$-dimensional entangling quantum system that sequentially interacts with qubits on a 1D semi-infinite conveyor belt. In each step $k$, the entangling unit can interact with two of these qubits, labelled $k$ and $k+N$, such that qubits $k+1,\dots k+N$ represent a quantum memory queue. (b) Physical interpretation: Wrapping the tape around a cylinder with the proper circumference, the entangling unit interacts with two qubits that are neighboring along the introduced vertical dimension. As time progresses the entangling unit moves along the tape creating a 2D entanglement structure. 
}
\label{fig1}
\end{figure}

To illustrate the main idea, we first consider a simplified setup  independent of any specific experimental implementation, where  a small quantum system \Q  is sequentially interacting with a string of qubits moving on a conveyor belt. In each discrete time step, \Q may write quantum states on initially uncorrelated qubits by unitary evolution and generate an output state.
If \Q interacts with one qubit (at position $k$) at each time $k$, it ``carries'' correlations from one qubit to another, thereby generating entanglement among them \cite{Schon:2005fk}.
In this approach, the size of \Q defines the information capacity and limits the maximum entanglement. 
In fact, the resulting quantum states can be exactly represented by so called matrix product state \cite{Schon:2005fk}, which naturally appear in one dimensional many-body systems \cite{Fannes:1992vq,White:1992zz,Vidal:2003gb,Orus:2014ja}.

The key idea of the present work is to  allow \Q to interact with qubits repeatedly and non-locally such that the information stored in step $k$ is also available in step $k+N$. 
This is achieved by additional interactions between \Q and qubit $k+N$ in step $k$  [Fig.~1(a)]. Such a storage and retrieval of information effectively realizes a  quantum queue memory of size $N$.
Owing to this memory, the resultant output state in general exhibits a qualitatively different entanglement structure. 
To visualize it, let us imagine rearranging the qubits such that the string winds around a cylinder with circumference $N$ as in Fig.~\ref{fig1}(b).
We interpret the new geometry as 2D square lattice with shifted periodic boundary conditions.
In this picture, \Q can create correlations between neighboring qubits not only in horizontal direction ($k$ and $k\pm1$) in subsequent steps, but also in vertical direction ($k$ and $k\pm N$) via simultaneous interactions in each turn of the protocol.
In particular, assuming that the qubits are initialized in the state $\ket{0}$, the unitary time evolution in each time step $k$ reduces to a map 
\begin{align}\label{eq1}
\hat{U}[k] = \sum_{i,a,b,c,d} U[k]_{a,b,c,d}^{i} \ket{i,a,b}\bra{c,0,d},
\end{align}
where $\ket{i,a,b} \equiv \ket{i}_k\ket{a}_{k+N}\ket{b}_\mathcal{Q}$ denotes a state with $k$-th and $k+N$-th qubits and $\mathcal{Q}$ in states $i$, $a$, $b$, respectively.
Repeated application of such maps produces a quantum state characterized by a 2D network of tensors $U[k]_{a,b,c,d}^{i}$.
This kind of representation corresponds to the description of 2D many-body systems within the framework of PEPS \cite{Verstraete:2008ko,Schuch:2007fm,Banuls:2008ch}, which include resources for measurement based quantum computation (MBQC) as well as topologically ordered systems.

 \begin{figure}[t]
\includegraphics[width=0.5\textwidth]{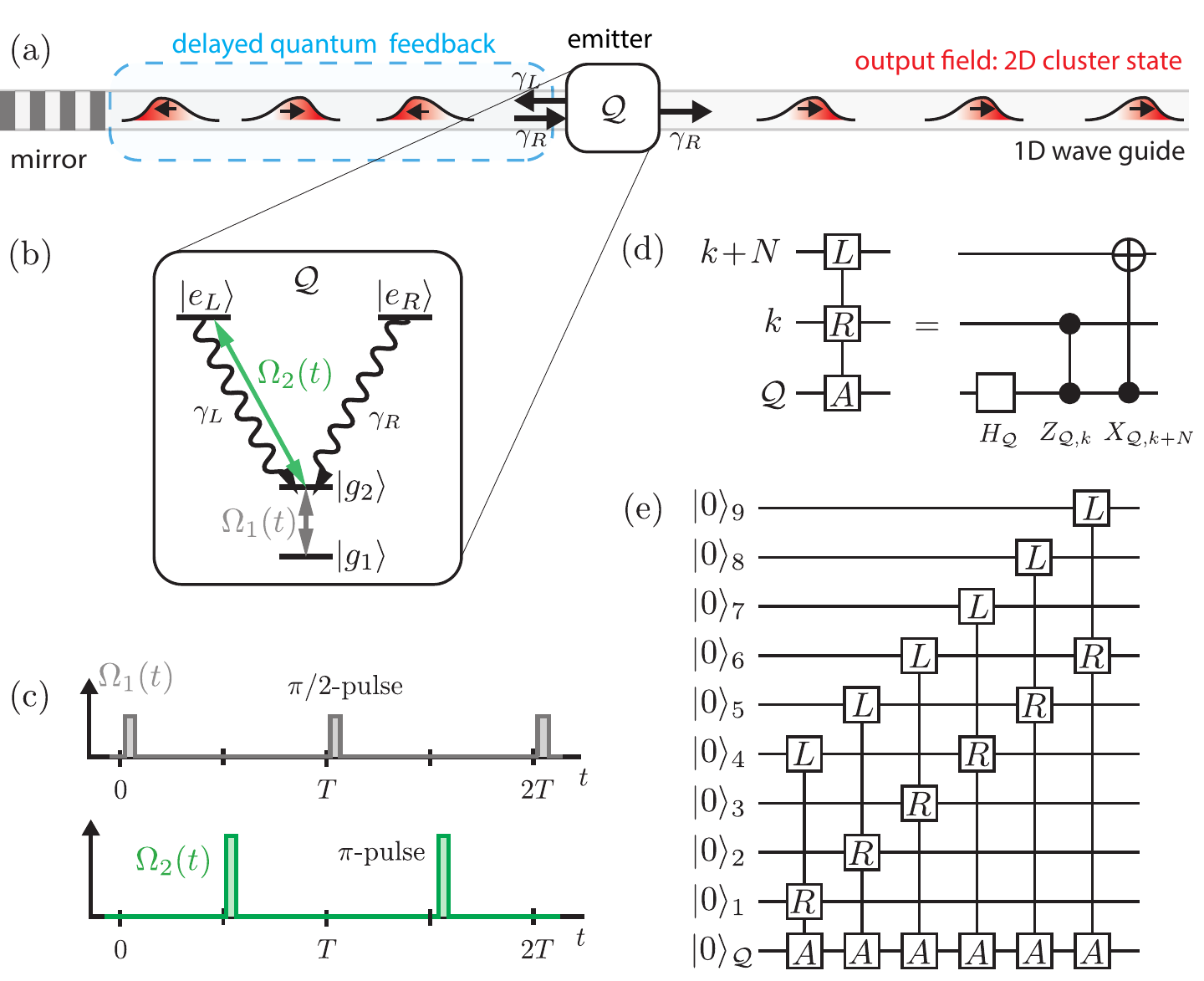}
\caption{Physical realization of the protocol to generate the photonic 2D cluster state. (a) A quantum emitter is coupled to a 1D wave-guide that is terminated on one side by a (distant) mirror. With the displayed atomic level structure (b) and periodic, pulsed coherent drive (c) the created photons form a 2D cluster state. The  corresponding circuit diagram is shown in (d,e). Each turn $k=1,2,\dots$ realized a unitary involving the atom ($A$), and two qubits, $k$ and $k+N$ [realized by a right ($R$) and left ($L$) moving photon respectively]. It consists of a Hadamard gate $\hat H_\mc{Q}$, and a controlled phase gate $\hat Z_{\mc{Q},k}$ between the atom an the photon qubit $k$ returning from the delay line, followed by a controlled-NOT gate $\hat X_{\mc{Q},k+N}$.}
\label{fig2}
\end{figure}

\textit{2D photonic cluster state.---} 
We now analyze this scheme in detail starting with the example of the 2D cluster state with a concrete physical realization. Specifically,  our scheme can be directly implemented in nanophotonic experiments.
We consider an individual quantum emitter (e.g.~an atom) representing $\mc{Q}$, coupled to an one-dimensional semi-infinite waveguide [Fig.~\ref{fig2}(a)]. This atom has two metastable  states $\ket{g_1}\equiv\ket{0}_\mc{Q}$ and $\ket{g_2}\equiv \ket{1}_\mc{Q}$, which can be coherently manipulated by a classical  field $\Omega_1(t)$ [Fig.~\ref{fig2}(b)]. 
The state $\ket{g_2}$ can be excited to state $\ket{e_L}$ using a laser with Rabi frequency $\Omega_2(t)$. Following each excitation the atom will decay to $\ket{g_2}$ emitting a photon into the waveguide. 
Qubits are encoded via the absence ($\ket{0}_k$) or presence ($\ket{1}_k$) of a photon during the time interval  $k$.
For now, we assume that the atom-photon coupling is chiral \cite{Lodahl:2017bz} such that all these 
photons are emitted unidirectionally by the atom into the waveguide, e.g. to the left in Fig.~\ref{fig2}(a).
The waveguide is terminated by a mirror located at a distance $L$ from the atom.
Finally, another excited state $\ket{e_R}$, degenerate with $\ket{e_L}$, couples to the right moving photons reflected from the mirror \cite{Guimond:2016kf}. We denote the corresponding decay rates by $\gamma_L$ and $\gamma_R$ [Fig.~\ref{fig2}(b)]. We note that alternative implementations that do not require chiral atom-photon interactions are also possible, as discussed below.

Our protocol starts by first generating 1D cluster states of left-propagating photons \cite{Lindner:2009eu}.
To this end, the atom is initially prepared in the state $\ket{0}_\mc{Q}$.
Then  a rapid $\pi/2$-pulse is applied on the atomic qubit, followed by a $\pi$-pulse on the $\ket{g_2} \rightarrow \ket{e_L}$ transition [Fig.~\ref{fig2}(c)].
The subsequent decay from state $\ket{e_L}$ to $\ket{g_2}$ results in entanglement between the atom and the emitted photon, i.e. $\ket{0}_\mc{Q}\ket{0}_1+\ket{1}_\mc{Q}\ket{1}_1$. 
When this pulse sequence is repeated for $n$ times, one can show that this protocol leads to a train of photonic qubits in the form of 1D cluster state. 
We note that an analogous scheme has been already demonstrated in an experiment using a quantum dot \cite{Schwartz:2016dj}, following a proposal by Lindner and Rudolph \cite{Lindner:2009eu}.

Interestingly, the 2D cluster state is generated from exactly the same sequence if we take into  account the effect of the mirror and the scattering of the right moving photons from the atom.
Each of the left-moving photons is reflected from the mirror and returns to the atom after a time delay $\tau=2L/c$, where $c$ denotes the speed of light. We are interested in the situation where this time delay is large so that the $k$-th photon interacts for the second time with the atom in between the two pulses of the $(k+N)$-th step of the protocol. This is  achieved, for example, by setting $\tau = (N-1/2)T$ where $T$ is the time duration of each time step [see Fig.\ref{fig2}(c)].
Crucially, when the atom is in the state $\ket{g_2}$, the right moving photon is resonantly coupled to the $\ket{g_2}\rightarrow \ket{e_R}$ transition, picking up a scattering phase shift of $\pi$ without any reflection \cite{Lodahl:2017bz}. In contrast, when the atom is in state $\ket{g_1}$, or the photon mode is empty, there is no interaction.
This process implements a controlled $\sigma^z$ gate 
\begin{align}\label{eqPhasegate}
\hat Z_{\mc{Q},k}=\ket{0}_\mc{Q}\bra{0}\otimes \mathbb{1}_k+\ket{1}_\mc{Q}\bra{1}\otimes \sigma_k^z
\end{align}
and entangles the atom and the $k$-th photon.
In turn, the subsequently generated $k+N$-th photon inherits this entanglement, thereby giving rises to the 2D structure described above.

Formally, the protocol can be interpreted as a sequential application of gates $\hat X_{\mc{Q},k+N} \hat Z_{\mc{Q},k}\hat H_\mc{Q}$, on the atom and (photonic) qubits $k$ and $k+N$, that are initially prepared in the trivial state $\ket{0}_\mc{Q}\bigotimes_k\ket{0}_k$ [Fig.~\ref{fig2}(d,e)]. Here $\hat H_\mc{Q}=\frac{1}{\sqrt{2}}(\sigma_\mc{Q}^{z}+\sigma_\mc{Q}^{x})$ is a Hadamard rotation of the atom and $\hat X_{\mc{Q},k}=\ket{0}_\mc{Q}\bra{0}\otimes \mathbb{1}_k+\ket{1}_\mc{Q}\bra{1}\otimes \sigma_k^x$ is a flip of the $k$-th qubit state, controlled by the atom. One can show [see supplementary material (SM)], that after $(M+1)\times N$ turns this gives exactly the 2D cluster state on a $M\times N$ square lattice with shifted periodic boundary conditions \footnote{Up to a trivial operation disentangling the atom.}:
\begin{align}\label{eq6}
\ket{\psi_{\mc{C}_{2D}}}=\lr{\prod_{k=1}^{N(M+1)}\hat X_{\mc{Q},k+N}\hat  Z_{\mc{Q},k} \hat H_\mc{Q} }\ket{0}_\mc{Q}\bigotimes_k\ket{0}_k.
\end{align}

\begin{figure}[t]
\includegraphics[width=0.5\textwidth]{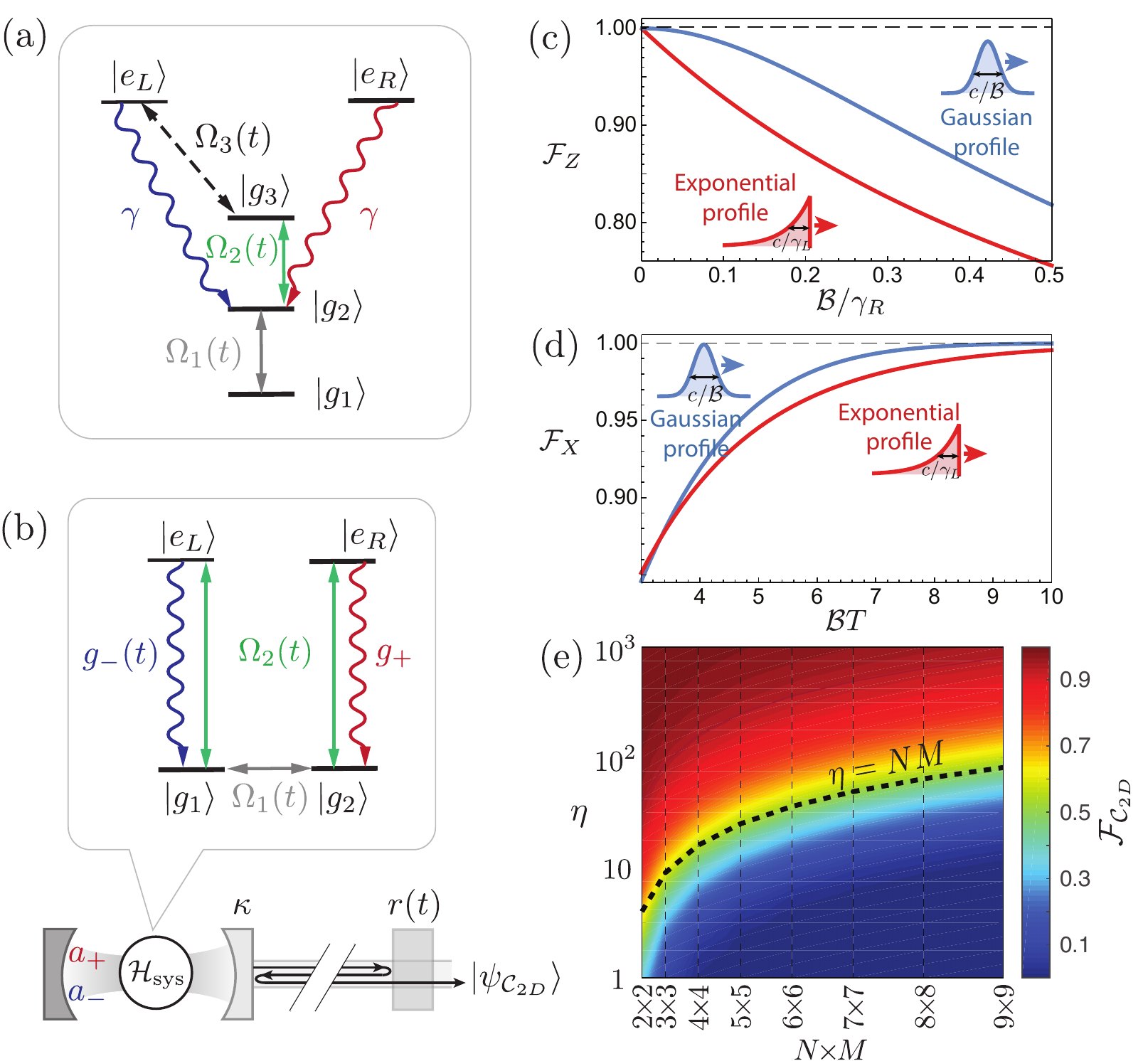}
\caption{Modified setups and effects of imperfections. (a) Modified setup to shape the photon wave-packet. In the limit $|\Omega_3(t)|\ll \gamma$ an adiabatic elimination of the excited state $\ket{e_L}$ leads to the same effective dynamics as in Fig.~\ref{fig2}(b), but with a renormalized and dynamically controllable, effective decay rate $\gamma_{L}(t)=|\Omega_3(t)|^2/(2\gamma)$. 
(b)  Modified setup to generate the 2D cluster state with photon qubits encoded in polarization degrees of freedom (i.e without the requirement of chiral coupling). (c) Fidelity of the controlled phase gate for a photon with Lorentzian (red) or Gaussian (blue) spectrum. (d) Analogously, fidelity of photon generation as a function of the period $T$ of a single step in the protocol. (e) Effect of photon loss: Fidelity of the generated state $\rho$ (with $\eta_L=\eta_R=\eta$) and the 2D cluster states of size $N\times M$, $\mathcal{F}_{\mc{C}_{2D}}=\sqrt{\bra{\psi_{\mc{C}_{2D}}}\rho\ket{\psi_{\mc{C}_{2D}}}}$.
}
\label{fig3}
\end{figure}

Before proceeding, we examine the necessary conditions for the implementation of our protocol.
First, photons generated in this pulsed scheme have a finite bandwidth $\mc{B}$. 
In order to realize the controlled phase gate \eqref{eqPhasegate} this bandwidth must be small, i.e., $\mc{B}\ll\gamma_R$ \cite{Lodahl:2017bz}; otherwise, the scattering by the atom does not only imprint a phase but also distort the photon wavepacket, reducing the gate fidelity, $\mc{F}_Z$ [Fig.~\ref{fig3}(c) and SM]. 
Narrow-bandwidth photons, and correspondingly, high fidelity gates, can be obtained by shaping the photon wavepacket, for example, by exciting the atom via a third stable level $\ket{g_3}$ in a Raman-type configuration [Fig.~\ref{fig3}(a)].  
In particular, the gate fidelity can be significantly increased by shaping the temporal profiles \cite{Pechal:2014db}, i.e. eliminating the error to first order in  $\mc{B}/\gamma_R$ [Fig.~\ref{fig3}(c) and SM]. 
Moreover, time symmetric photons simplify the measurement of the qubits in an arbitrary basis, since they can be perfectly absorbed by a second atom acting as the measurement device \cite{Cirac:1997is}. Such techniques allow the full implementation of MBQC using our protocol \cite{Rausssendorf:2001js}.

Second, the information capacity of the queue memory is bounded by the number of photons in the delay line. 
In order to well distinguish two consecutive photons, one can only generate them at a rate $1/T\ll \mc{B}$ [see Fig.~\ref{fig3}].
Therefore, the requirement for narrow-bandwidth photons is competing with the effective size of the achievable memory, $N$. 
Thus, high fidelity implementation requires the hierarchy
\begin{align}
N\sim \tau/T\ll \tau\mc{B}\ll\gamma_R \tau.
\end{align} 
Note that the quantum memory lifetime can be dramatically enhanced e.g. via a dispersive slow-light medium \cite{Lukin:2003ct,Fleischhauer:2005ti}.

Finally, apart from this fundamental considerations, experimental imperfections will eventually limit the achievable size of the cluster state. One of the most important challenges is the photon loss, often quantified by the so-called cooperativity $\eta_j=\gamma_j/\Gamma_j$ ($j=L,R$). Here $\Gamma_j$ denotes the effective rate of photon loss arising from emission into unguided modes (from state $\ket{e_j}$) and amplitude attenuation in the waveguide. Large, high-fidelity cluster states can be obtained in the regime $\eta_j\gg 1$, where the achievable system sizes scale as  $NM\lesssim (1/\eta_L+2/\eta_R)^{-1}$ [Fig.~\ref{fig3}(e)].  High cooperativities have been demonstrated  in nanophotonic experiments with neutral atoms and solid-state emitters \cite{Tiecke:vw}.

We also note that our protocol can be adapted for settings that do not have chiral couplings. For example, when an atom is coupled to a waveguide via a one-sided cavity \cite{Tiecke:vw,Reiserer:2014hf}, the delayed feedback can  be introduced by a distant, switchable mirror [Fig.~\ref{fig3}(b)]. Proper control of the mirror can ensure that each generated photon interacts exactly twice with the atom before it leaves at the output port, as required for our protocol. Moreover, in such a setting it is possible to encode qubit states in photon polarizations rather than number degrees of freedom  [see Fig.~\ref{fig3}(b)], allowing the detection of photon loss errors.

\textit{Generalizations and Outlook}.--- 
Apart from applications in quantum computing, our scheme can be harnessed to study strongly correlated quantum systems. Output fields of a quantum emitter can be used as a variational class to search for ground states of Hamiltonians of interest. While this has been discussed previously for 1D problems \cite{Haegeman:2010bc,Eichler:2015if}, where the generated states are limited to MPS and can be simulated classically \cite{Vidal:2008bk},  a 2D tensor network is qualitatively different from a MPS, since an exact contraction, e.g. to calculate correlation functions, is in general intractable on classical computers  \cite{Schuch:2007fm}. Moreover, our scheme allows study of many-body phenomena that are present only in dimensions larger than one, such as topologically non-trivial phases. 

\begin{figure}[t!!]
\includegraphics[width=0.45\textwidth]{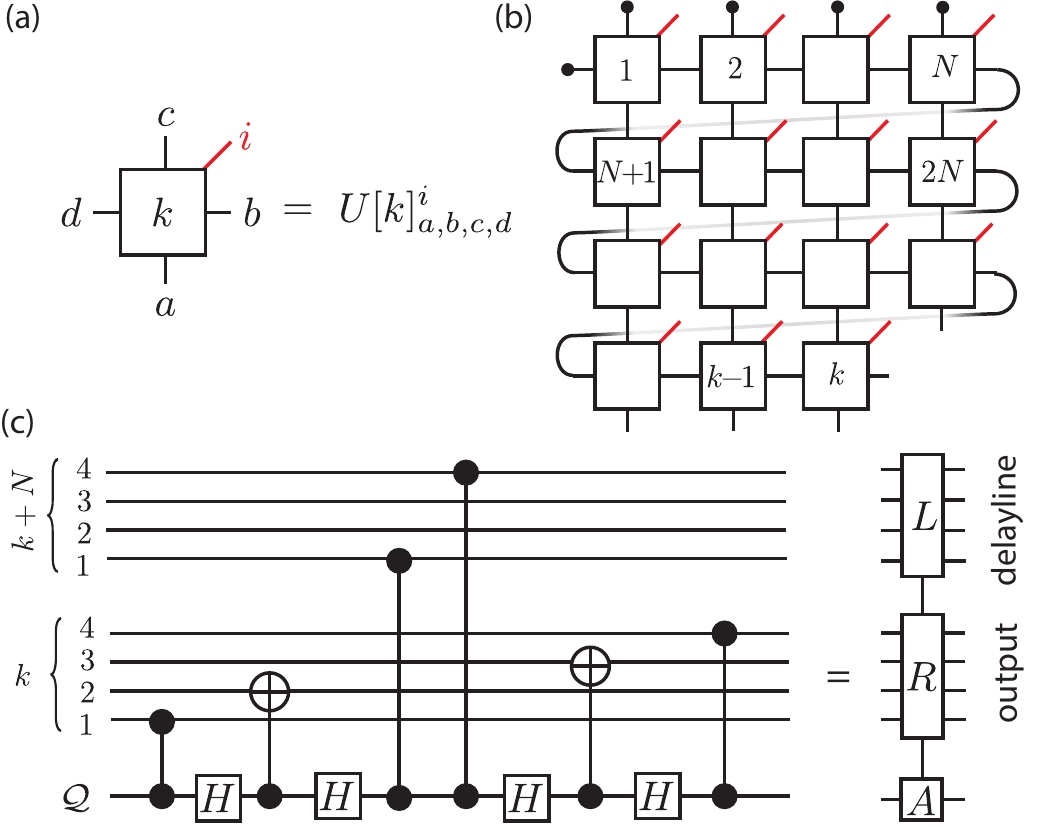}
\caption{Tensor network representation of the generated states. (a) Graphical representation of the tensors $U[k]$.  (b) Representation of the state \eqref{eq5} in terms of the tensors given in (a). Connected lines indicate contractions, red open lines denote physical indices corresponding to the states of qubits in the output. In the top row the tensors are contracted with the initial state of the \Q and the first $N$ qubits in the string, as indicated by the  black circles. The open legs at the bottom correspond to the state of the qubits in the memory queue after step $k$, and the last open line on the bottom right, corresponds to the state of $\mc{Q}$. (c) Circuit model that gives the tensor \eqref{eq8} (up to local unitary operations) for an uncorrelated initial state in which \Q and qubits $k_1,k_4$ ($k=1,2,\dots$) are in state $(\ket{0}+\ket{1})/\sqrt{2}$ and all other qubits in $\ket{0}$ [cf.~Fig.~\ref{fig2}(d,e)]}
\label{fig4}
\end{figure}

Indeed, the class of states that can be created in our approach can be completely characterized.
More specifically, we are interested in the structure of the wavefunction $\ket{\Psi(k)}$, describing \Q and the string of qubits, after $k$ steps:
\begin{align}\label{eq5}
\ket{\Psi(k)}&=\hat U[k]\ket{\Psi(k-1)}=\prod_{j=1}^k\hat U[j] \ket{\Psi(0)},
\end{align}
where the unitary evolution $\hat U[j]$ acts only on \Q and qubits $j$ and $j+N$ and, for concreteness, we choose $\ket{\Psi(0)}=\ket{0}_\mc{Q}\bigotimes_k\ket{0}_k$ \footnote{Note that we use a convention where the ordering in the product is defined via $\prod_{j=1}^k M_j=M_kM_{k-1}\dots M_1$.}. 
Given the initial state and the limited support of $\hat{U}$, the wavefunction is entirely specified by $U[j]_{a,b,c,d}^i$ (see eq.~\eqref{eq1}).
In particular, $U[j]_{a,b,c,d}^i$ can be understood as a rank-5 tensor; where the ``physical index'' $i$ denotes the state of the qubit $j$, the ``horizontal bonds'' $b$ and $d$ run over internal degrees of freedom of $\mc{Q}$,  and ``vertical bonds'' $a$ and $c$ enumerate the quantum states of input into and output from queue memory, respectively  [see Fig.~\ref{fig4}(a)]. 
Therefore, we can write $\ket{\Psi(k)}$ as
\begin{align}\label{eq3}
\ket{\Psi(k)}=\sum_{i_\mc{Q},\{i_j\}} \mc{C}(\{U[j]^{i_j}\})\ket{i_\mc{Q},i_1,i_2,\cdots},
\end{align}
where $\mc{C}(\dots)$ denotes the contraction of the 2D tensor networks in Fig.~\ref{fig4}(b) and $\ket{i_\mc{Q},i_1,i_2,\cdots}$ enumerates configurations of \Q and the string of qubits.
Since the tensors describing the state are given by the matrix elements of a unitary matrix, they have to satisfy 
\begin{align}\label{eq7}
\sum_{i,a,b}U^{i}_{a,b,c,d}(U^{i}_{a,b,c',d'})^\ast=\delta_{c,c'}\delta_{d,d'}.
\end{align}
Physically, this isometric condition reflects the deterministic and sequential nature of the protocol --- at a given step $k$ the state of qubits $j<k$ can not be changed. 
Every tensor network that can be brought into a form respecting \eqref{eq7} can be constructed in our approach. 
Per construction we showed above that this includes universal resources for MBQC. 

Interestingly, this class of states includes exotic states with topological order such as string-net states \cite{Buerschaper:2009fi} or the ground state of Kitaev's Hamiltonian \cite{Kitaev:2003jw}. In the latter case the ground state can be represented by a 2D network of translationally invariant tensors \cite{Orus:2014ja,Schuch:2010jp,Schuch:2012cq}
\begin{align}\label{eq8}
U^{i_1,i_2,i_3,i_4}_{a,b,c,d}=\delta_{a+b,i_1}\delta_{d+a,i_2}\delta_{c+d,i_3}\delta_{b+c,i_4}/4,
\end{align}
where $i_1, \dots, i_4$ denote the states of four physical qubits in a unit cell and $\{a, b, c, d\}$ run over the bond dimension 2.
This tensor explicitly satisfies \eqref{eq7} and can immediately be translated into a protocol
similar to the one for generating the 2D cluster state.
 In Fig. \ref{fig4}(c) we give an explicit circuit representation of the stroboscopic step $\hat U$, where we only utilize gates that are accessible in photonic systems discussed above, such as 
 controlled single photon generation $\hat X_{\mc{Q},k}$ and atom-photon phase gates $\hat Z_{\mc{Q},k}$. The four physical qubits of each unit cell are in this case represented by four sequentially generated photons in each  step.

Our work can be extended in several ways. In analogy to continuous MPS in 1D \cite{Verstraete:2010bf}, our protocol can be adapted to use continuos driving fields, which results in hybrid continuous-discrete 2D tensor network states. Moreover, adding multiple delay lines allows \Q to interact with more than two qubits, and thus provides tensor networks in higher dimensions. This is of relevance for fault tolerant implementations of MBQC using 3D cluster states \cite{Briegel:2009gg,Raussendorf:2006je}. A promising extension of our nanophotonic protocol is implementation of MBQC that can tolerate up to 50\% of counterfactual errors due to photon loss \cite{Varnava:2006jd,Dawson:2006eo}. Finally, besides nanophotonic setups our `single-atom quantum computer' can be implemented with microwave photons \cite{vanLoo:2013df,Hoi:2015fh}, phonons \cite{Ramos:2014ut}, surface-acoustic waves \cite{Gustafsson:2014gu,Guo:2016ui} or moving tapes of spin qubits \cite{Benito:2016cw}.

\begin{acknowledgements}
We thank J.~I.~Cirac, J.~Haegeman, L.~Jiang, N.~Schuch and F.~Verstraete for useful discussions. This work was supported through NSF, CUA, AFOSR Muri and the Vannevar Bush Faculty Fellowship. H.~P. is supported by the NSF through a grant for the Institute for Theoretical Atomic, Molecular, and Optical Physics at Harvard University and the Smithsonian Astrophysical Observatory. S.~C. acknowledges the support from Kwanjeong Educational Foundation. Work at Innsbruck is supported by SFB FOQUS and ERC Synergy Grant UQUAM.
\end{acknowledgements}

\bibliographystyle{apsrev4-1.bst}
\bibliography{HannesBib}

\onecolumngrid
\newpage
{
\center \bf \large 
Supplemental Material for: \\
Photonic tensor networks produced by a single quantum emitter\vspace*{0.1cm}\\ 
\vspace*{0.0cm}
}
\begin{center}
Hannes Pichler$^{1,2}$, Soonwon Choi$^{2}$, Peter Zoller$^{3,4}$, and Mikhail D. Lukin $^{2}$\\
\vspace*{0.15cm}
\small{\textit{$^1$ITAMP, Harvard-Smithsonian Center for Astrophysics, Cambridge, MA 02138, USA\\
$^2$Physics Department, Harvard University, Cambridge, Massachusetts 02138, USA\\
$^3$Institute for Theoretical Physics, University of Innsbruck, A-6020
Innsbruck, Austria\\
$^4$ Institute for Quantum Optics and Quantum Information of the Austrian
Academy of Sciences, A-6020 Innsbruck, Austria\\}}
\vspace*{0.25cm}
\end{center}

\twocolumngrid

\section{2D cluster state representation} 
In this section we prove that $\ket{\psi_{\mc{C}_{2D}}}$ in Eq.~\eqref{eq6} represents the 2D cluster state.
To this end we start first with the representation of a the 1D cluster state on $K+1$ qubits:
\begin{align}
\ket{\psi_{\mc{C}_{1D}}}=\prod_{k=1}^{K} Z_{k,k+1}\ket{+}^{\otimes {K+1}}.
\end{align}
Using the swap operator $S_{i,j}$ that exchanges the quantum states of qubits $i$ and $j$ we can rewrite the 1D cluster state as
\begin{align}
\ket{\psi_{\mc{C}_{1D}}}=\lr{\prod_{k=1}^{K} S_{\mc{Q},k}Z_{\mc{Q},k} }\ket{+}_\mc{Q}\bigotimes_{k=1}^{K}\ket{+}_k
\end{align}
where we used the relation $Z_{b,c}=S_{a,b}Z_{a,c}S_{a,b}$, and identified the $K+1$th qubit with the ancilla $\mc{Q}$. Note that throughout this paper we use a convention where the ordering in the product is defined via $\prod_{j=1}^k M_j=M_kM_{k-1}\dots M_1$.
We now use the relation  
\begin{align}
S_{\mc{Q},v}Z_{\mc{Q},v}\ket{\psi}_\mc{Q}\otimes\ket{+}_v= H_\mc{Q} X_{\mc{Q},v} H_v \ket{\psi}_\mc{Q}\otimes\ket{+}_v,
\end{align}
where $H_x$ is the Hadamard gate acting on qubit $x$. 
We note that this equality is not an operator identity but a property of states of the form $\ket{\psi}_\mc{Q}\otimes\ket{+}_v$, where the qubit $v$ must be in the state $\ket{+}_v$ while the ancillary system $\ket{\psi}_\mc{Q}$ may be in an arbitrary state, potentially entangled to other systems. 
Using these relations, our state can be written as 
\begin{align}
\ket{\psi_{\mc{C}_{1D}}}=\lr{\prod_{k=1}^K H_\mc{Q} X_{\mc{Q},k}} \ket{+}_\mc{Q} \bigotimes_{k=1}^K\ket{0}_k.
\end{align}
We note that this representation of the 1D cluster state has been also used in Ref.~\cite{Lindner:2009eu}.

We now proceed to the construction of the 2D cluster state.
From its definition, the 2D cluster state can be obtained from the 1D cluster state above by introducing additional entanglement (via phase gates) between qubits $k$ and $k+N$~\cite{Rausssendorf:2001js}. This gives a 2D cluster state on a square lattice with shifted periodic boundary conditions, where the extend of the (shifted) periodic direction is set by $N$. 
\begin{align}
\ket{\psi_{\mc{C}_{2D}}}&=\lr{\prod_{k=N+1}^{K}Z_{k,k-N}}\lr{\prod_{k=1}^K H_\mc{Q} X_{\mc{Q},k}} \ket{+}_\mc{Q} \bigotimes_{k}\ket{0}_k\no\\
&=\lr{\prod_{k=1}^K H_\mc{Q} Z_{k,k-N}X_{\mc{Q},k}} \ket{+}_\mc{Q} \bigotimes_{k}\ket{0}_k.
\end{align}
In the second line we used the fact the $[Z_{j,j+N},X_{\mc{Q},k}]=0$ for $j>k$, and $Z_{k,j}\ket{0}_j=\mathbb{1}_k\otimes\mathbb{1}_{j}\ket{0}_j$.
Now we make use of the identity
\begin{align}
Z_{n,m}X_{\mc{Q},n}\ket{\psi}_{\mc{Q},m}\ket{0}_n=X_{\mc{Q},n}Z_{\mc{Q},m}\ket{\psi}_{\mc{Q},m}\ket{0}_n,
\end{align}
where $\ket{\psi}_{\mc{Q},m}$ is an arbitrary state of $\mc{Q}$ and every qubit including $m \neq n$ but not $n$.
Again, this relation is not an operator identity, but a property of a state where qubit $n$ is in the separable state $\ket{0}_n$. 
Using this identity we arrive at 
\begin{align}
\ket{\psi_{\mc{C}_{2D}}}=\lr{\prod_{k=1}^{K} H_\mc{Q} X_{\mc{Q},k} Z_{\mc{Q},k-N}}\ket{+}_\mc{Q}\bigotimes_k\ket{0}_k
\end{align}
which is (up to a shift of the index and a rotation of $\mc{Q}$) exactly our protocol given in eq.~\eqref{eq6}.

\section{Imperfections}
\subsection{Phase gate fidelity due to finite bandwidth of photons.}
Without shaping the wave packets of the emitted photons, each photon produced in a single step has a Lorentzian spectral profile, whose temporal profile is 
\begin{align}\label{unshaped_photon}
f(t)=\sqrt{\gamma_L}e^{-\gamma_L t/2}\Theta(t)=i \int d\omega\,\frac{2\pi}{\sqrt{\gamma_L}}\frac{1}{\omega+i\gamma_L/2}e^{-i\omega t}
\end{align}
where we chose a normalization $\int dt|f(t)|^2=1$. 
The scattering phase shift for the chiral forward scattering (if the atom is in state $\ket{g_2}$) is determined by the transmission:
\begin{align}\label{eqtransmission}
t(\omega)=\frac{\omega-i\gamma_R/2}{\omega+i\gamma_R/2}
\end{align}
such that the wave packet $f(t)$ transforms into
\begin{align}
\tilde f(t)&=i \int d\omega\,\frac{2\pi}{\sqrt{\gamma_L}}\frac{\omega-i\gamma_R/2}{\omega+i\gamma_R/2}\frac{1}{\omega+i\gamma_L/2}e^{-i\omega t}\\
&=-\sqrt{\gamma_L}\lr{\frac{\gamma_R+\gamma_L}{\gamma_R-\gamma_L}e^{-\gamma_L t/2}-2\frac{\gamma_R}{\gamma_{R}-\gamma_L}e^{-\gamma_R t/2}}\Theta(t)
\end{align}
It is straightforward to calculate the overlap
\begin{align}
\int dt f^\ast(t) \tilde f (t)=-\frac{1-\gamma_L/\gamma_R}{1+\gamma_{L}/\gamma_R} 
\end{align}
With this we obtain the fidelity \cite{Nielsen:2002ks} of the controlled phase gate:
\begin{align}
\mathcal{F}_Z=\frac{2 \frac{\gamma_l}{\gamma_{R}} (\frac{\gamma_l}{\gamma_{R}}+3)+5}{5 (\frac{\gamma_l}{\gamma_{R}}+1)^2}=1-\frac{4}{5}\frac{\gamma_L}{\gamma_R}+\mc{O}\lr{\frac{\gamma_L^2}{\gamma_R^2}}.
\end{align}
 \subsubsection{Pulse shaping}
If we shape the coupling via $\gamma_L(t)=\frac{4 \Omega^2(t)}{\gamma}$ then $f(t)=
\sqrt{\gamma_L(t)} \exp\lr{-\int_0^t ds \gamma_L(s)/2} $.
Pulse shaping allows to create photon wave packets that are symmetric in time. Straightforward calculation shows that, for example,  the gaussian wave packet 
\begin{align}
f(t)=\sqrt{\mc{B}/\sqrt{\pi}}e^{-\mc{B}^2 (t-t_0)^2/2}
\end{align}
can be obtained by 
\begin{align}\label{shaped_decay}\gamma_L(t)=\frac{2\mc{B} e^{-\mc{B}^2 (t-t_0)^2}}{\sqrt{\pi}(1-{\rm erf}(\mc{B} (t-t_0)))}.
\end{align}
The corresponding fidelity of the phase gate can be calculated from
\begin{align}
\int dt f^\ast(t) \tilde f (t)&=1-\frac{\sqrt{\pi } e^{\frac{1}{4 x^2}}\lr{1- \text{erf}\left(\frac{1}{2 x}\right)}}{x}\no\\&=-1 + 4 x^2+\mc{O}(x^4)
\end{align}
with $x=\mc{B}/\gamma_R$, and the gaussian error function $\textrm{erf}(z)=\frac{2}{\sqrt{\pi}}\int_0^z dt e^{-t^2}$. This gives the fidelity $\mc{F}_Z=1-\frac{8}{5}x^2+\mathcal{O}(x^4)$  \cite{Nielsen:2002ks}.
We note that the linear order in $x$ vanishes in this expression unlike in the previous a case without the shaping of wave packet.
This is a consequence of temporally symmetric wave packet, which can significantly improve the fidelity. 

\subsection{Fidelity of the controlled not gate}
In the proposed implementation to create the 2D cluster state, the gate $\hat X_{\mathcal{Q},k+N}$ is realized by emission of a photon associated with the transition $\ket{e_L} \rightarrow \ket{g_2}$ during the second half of each timestep with period $T$. In order for the gate $\hat X_{\mathcal{Q},k+N}$ to work, we thus require $T/2$ to be much larger than the temporal extend of the emitted photon; otherwise, the next step in our protocol would proceed even before the gate $\hat{X}_{\mathcal{Q},k_N}$ is completed, leading to an error. 

This gate fidelity can computed from the quantity 
\begin{align}
\epsilon=\int_{t_0}^{t_0+T/2} dt \gamma_L(t) \exp\lr{-\int_0^t ds \gamma_L(s)}
\end{align}
via $\mc{F}_X=1-\frac{2}{3}\epsilon+\frac{1}{6}\epsilon^2$.
Without shaping the photon wave packet, i.e., with $\gamma_L(t)=\gamma_L$ as in eq.~\eqref{unshaped_photon} one gets $\epsilon=e^{-\gamma_L T/2}$, while for the pulse shaped photon \eqref{shaped_decay}  (with $t_0=-T/4$) we find $\epsilon=1-\sqrt{2}\frac{\textrm{erf}(\mc{B} T/4)}{\sqrt{1+\textrm{erf}(\mc{B} T/4)}}$. For large $x=\mc{B} T/4 \gg1$ we have $\epsilon\rightarrow e^{-x^2}/(\sqrt{\pi}x)$. In both cases the gate fidelity approaches 1 exponentially, but in the case of a shaped photon wave packet this approach is again faster. 

\end{document}